\documentstyle[11pt]{article}
\begin{document}


\thispagestyle{empty}

\newcommand{\al}{\alpha}
\newcommand{\bet}{\beta}
\newcommand{\ga}{\gamma}
\newcommand{\del}{\delta}
\newcommand{\ep}{\epsilon}
\newcommand{\epx}{\varepsilon}
\newcommand{\ze}{\zeta}
\newcommand{\th}{\theta}
\newcommand{\thx}{\vartheta}
\newcommand{\io}{\iota}
\newcommand{\la}{\lambda}
\newcommand{\ka}{\kappa}
\newcommand{\pix}{\varpi}
\newcommand{\rhx}{\varrho}
\newcommand{\si}{\sigma}
\newcommand{\six}{\varsigma}
\newcommand{\yp}{\upsilon}
\newcommand{\om}{\omega}
\newcommand{\phx}{\varphi}
\newcommand{\Ga}{\Gamma}
\newcommand{\De}{\Delta}
\newcommand{\Th}{\Theta}
\newcommand{\La}{\Lambda}
\newcommand{\Si}{\Sigma}
\newcommand{\Yp}{\Upsilon}
\newcommand{\Om}{\Omega}


\newcommand{\be}{\begin{eqnarray}}
\newcommand{\ee}{\end{eqnarray}}
\newcommand{\jt}{\tilde{J}}
\newcommand{\Ra}{\Rightarrow}
\newcommand{\lra}{\longrightarrow}
\newcommand{\Llra}{\Longleftrightarrow}
\newcommand{\pr}{\partial}
\newcommand{\ti}{\tilde}
\newcommand{\ng}{|0\rangle_{gh}}
\newcommand{\pj}{\prod J}
\newcommand{\pjt}{\prod\tilde{J}}
\newcommand{\prb}{\prod b}
\newcommand{\prc}{\prod c}
\newcommand{\bft}{|\tilde{\phi}>}
\newcommand{\bfj}{|\phi>}
\newcommand{\lan}{\langle}
\newcommand{\ran}{\rangle}
\newcommand{\bz}{\bar{z}}
\newcommand{\bJ}{\bar{J}}
\newcommand{\vacr}{|0\rangle}
\newcommand{\vacl}{\langle 0|}
\newcommand{\IFF}{\Longleftrightarrow}
\newcommand{\phr}{|phys\ran}
\newcommand{\phl}{\lan phys|}
\newcommand{\rar}{\rightarrow}
\newcommand{\da}{\downarrow}


\newcommand{\ind}{\indent}
\newcommand{\np}{\newpage}
\newcommand{\hs}{\hspace}
\newcommand{\vs}{\vspace}
\newcommand{\nl}{\newline}
\newcommand{\nn}{\nonumber}
\newcommand{\lef}{\left}
\newcommand{\rig}{\right}
\newcommand{\fra}{\twelvefrakh}
\newcommand{\Bb}{\twelvemsb}
\newcommand{\bT}{\bar{T}(\bz)}
\newcommand{\bD}{\bar{D}}
\newcommand{\bth}{\bar{\th}}
\newcommand{\bla}{\bar{\la}}
\newcommand{\bga}{\bar{\ga}}
\newcommand{\bbet}{\bar{\bet}}

\newcommand{\bZ}{\bar{Z}}
\newcommand{\bA}{\bar{A}}
\newcommand{\Ai}{A^{-1}}
\newcommand{\Gi}{G^{-1}}
\newcommand{\Hi}{H^{-1}}
\newcommand{\gi}{g^{-1}}
\newcommand{\hi}{h^{-1}}
\newcommand{\hk}{\hat{k}}
\newcommand{\hkt}{\hat{\tilde{k}}}
\newcommand{\hJ}{\hat{J}}

\newcommand{\bC}{\bar{C}}
\newcommand{\bc}{\bar{c}}
\newcommand{\bB}{\bar{B}}
\newcommand{\bb}{\bar{b}}

\newcommand{\hJt}{\hat{\ti{J}}}
\newcommand{\dagg}{^{\dagger}}
\newcommand{\qd}{\dot{q}}
\newcommand{\cP}{{\cal P}}
\newcommand{\hg}{\hat{g}}
\newcommand{\tb}{\bar{\th}}
\newcommand{\hh}{\hat{h}}
\newcommand{\hpg}{\hat{g}^\prime}
\newcommand{\htg}{\tilde{\hat{g}}^\prime}
\newcommand{\pri}{^\prime}
\newcommand{\lap}{\la^\prime}
\newcommand{\rhop}{\rho^\prime}
\newcommand{\Dgp}{\Delta_{g^\prime}^+}
\newcommand{\Dg}{\Delta_g^+}
\newcommand{\Pro}{\prod_{n=1}^\infty (1-q^n)}
\newcommand{\Pg}{P^+_{\hg}}
\newcommand{\Pgp}{P^+_{\hg\pri}}
\newcommand{\hmu}{\hat{\mu}}
\newcommand{\hnu}{\hat{\nu}}
\newcommand{\hrho}{\hat{\rho}}
\newcommand{\tla}{\tilde{\la}}
\newcommand{\btla}{\bar{\tilde{\la}}}
\newcommand{\gp}{g^\prime}
\newcommand{\pp}{\prime\prime}


\newcommand{\NPB}[1]{Nucl. Phys. B  {#1} }
\newcommand{\IJMPA}[1]{Int. J. of Mod. Phys. A  {#1} }
\newcommand{\PLB}[1]{Phys. Lett. B {#1} }
\vs*{-15mm}
\begin{flushright}
G\"oteborg ITP 95-23 \\
September 1995 \\
\vs{8mm}
\end{flushright}

\begin{center}

{\huge{Gauged supersymmetric WZNW model \\
\vspace{3mm}
using the BRST approach}} \\
\vspace{10 mm}
{\Large{Henric Rhedin}\footnote{hr@fy.chalmers.se}} \\
\vspace{4 mm}

Institute of Theoretical Physics \\
G\"oteborg University and \\
Chalmers University of Technology \\
S-412 96 G\"oteborg, Sweden \\

\vs{10mm}

{\bf{Abstract}} \end{center}
\begin{quotation}

\noindent We consider the supersymmetric WZNW model gauged
in a manifestly supersymmetric way. We find the BRST charge
and the necessary condition for nilpotency. In the BRST
framework the model proves to be a Lagrangian formulation
of the supersymmetric coset construction, known as
the N=1 Kazama-Suzuki coset construction.

\end{quotation}

\np
\setcounter{page}{1}

\section{Introduction}

In the context of conformal field theory, much effort has been
devoted to the WZNW model, especially gauged versions
thereof, which are generally believed to
describe all rational conformal field theories.
In the gauged version the degrees of freedom of a subalgebra h of the
full algebra g is removed by the procedure known as the gaugeing.
On the algebra level, the counterpart is the
coset conformal field theories known as the Goddard-Kent-Olive
construction \cite{GKO}. Those two approaches were connected when,
in the BRST framework, the gauged WZNW model was shown
to be the Lagrangian formulation of the algebraic approach \cite{KS,HR1}.

If we want to discuss realistic string theories, we would, however,
like to incorporate fermions. This can be done by supersymmetrizing
the WZNW model \cite{DKPR,AA}. The resulting action may
be rewritten as a bosonic WZNW model and free fermions.
In the algebraic approach, this may be formulated as a coset construction
\cite{KaSu} in much the same way as for the bosonic case
and we call this the Kazama-Suzuki coset construction. This construction
gives an orthogonal decomposition of the $N=1$ superconformal
algebra in the sense that it enables the definition of coset
superconformal generators which decouple from the subalgebra generators.
This was also discussed in \cite{GKO} for the special case where the
full diagonal algebra is taken as the subalgebra in the coset
construction i.e. (G$\otimes$G)/G models. This
provides with what is known as the minimal superconformal models.

Some time ago, a Lagrangian formulation of the minimal superconformal
model was found by Schnitzer \cite{Sc}. The starting point was the
supersymmetric WZNW model. It was, however, gauged only by bosonic degrees of
freedom
and after the bosons and the fermions were decoupled.
The resulting effective action is composed of the original bosonic
WZNW model, an auxiliary bosonic WZNW model resulting from the
gaugefields, the original fermions, and a fermionic ghost system.
This resulted in a superconformal algebra which closed only
in physical amplitudes. It was suggested
that this lack of closure was due to the fact that the supersymmetric
WZNW model had not been gauged in a manifestly supersymmetric way.

The manifestly supersymmetric gauging of the supersymmetric WZNW
model was discussed in \cite{Ts}. The effective action here consists of
two supersymmetric WZNW models, the original as well as an auxiliary
one originating from the gaugeing. Here it was also found that
the conformal anomaly of the effective action of the gauged
supersymmetric WZNW model coincided with that of the Kazama-Suzuki
coset construction.

We here discuss the correspondence between the manifestly
supersymmetrically gauged supersymmetric WZNW model and the
Kazama-Suzuki coset
construction. We approach this issue in the BRST framework.
To this end, we introduce a superconformal ghost system which
may be represented by a fermionic ghost system as well as a bosonic
ghost system. The BRST symmetry of the effective action is found
and the BRST charge is presented.

We proceed by showing that the generators of the $N=1$ superconformal
algebra of both the full algebra and the subalgebra are BRST invariant.
Furthermore, we find that the generators of the $N=1$ superconformal
algebra of the effective action are given by the Kazama-Suzuli coset
generators plus a BRST exact term.
We hence provide with a Lagrangian formulation of
the Kazama-Suzuki $N=1$ coset construction.

There exists, however, a different approach to the supersymmetric
gauged WZNW model. One may decouple the bosons and the fermions
and discuss the BRST symmetry of the resulting system. This gives
us another model which resembles the one in \cite{Sc}.
Specifically the two approaches yield different nilpotency conditions
for the BRST charges and this results in different conformal anomalies.
They, however, coincide for g=h. For g$\neq$h
the exact relation between those models is still unclear. \\

\section{Preliminaries}

We consider the supersymmetric WZNW model in
superspace
defined from the action \cite{DKPR,AA}
\be
S_{SWZNW}=\frac{\hk}{4\pi}\int{\rm d}Z{\rm d}\bar{Z}\lef(-D{\cal G}
\bD{\cal G}^{-1}+
\int{\rm d}t{\cal G}\frac{{\rm d}}{{\rm d}t}{\cal G}^{-1}
(D{\cal G}\bar{D}{\cal G}^{-1}+
\bar{D}{\cal G}D{\cal G}^{-1})\rig) \label{action}
\ee
in terms of the superfield ${\cal G}$ which may be expanded in components as
${\cal G}(Z,\bZ)=g(z,\bz)+\th\psi_1(z,\bz)+\bth\psi_2(z,\bz)+\th\bth F(z,\bz)$.
Here $g$ take values in a group G,
$\psi_1$ and $\psi_2$ are the components of a Majorana fermion, and
$F$ is an auxiliary field. Also we have
$D=\pr_{\th}+\th\pr_z, \ \bD=\pr_{\bth}+\bth\pr_{\bz}$.
The equations of motion have the form $\bD(D{\cal G}{\cal G}^{-1})=0$.
Written in components and using the equations of motions to
eliminate the auxiliary field $F$
the action may be put in the form of an ordinary bosonic WZNW
model defined over $g$ and free fermions $\la(z)=\psi_1\gi$ and
$\bar{\la}(\bz)=\gi\psi_2$.

Superconformal transformations are generated by
$K(Z)=\frac{1}{2}G(z)+\th L(z) \label{superconf}$
where $G(z)$ is the supersymmetry generator and $T(z)$ is the stress
tensor obeying an $N=1$ superconformal algebra.

The action is invariant under the superaffine transformations.
The conserved current corresponding to this symmetry (we only
bother to display the superholomorphic sector, the other
sector is completely analogous)
$j(Z)=-\frac{\hk}{2}D{\cal G}{\cal G}^{-1}$.
In components we have
$D{\cal G}{\cal G}^{-1}=\la+\th(\pr_zg\gi+\la\la)\equiv \la+\th\hJ$
where we have used that $\bla$ only depends on $\bz$.
We have the superaffine operator product expansions (OPE's)
\be
& &\hJ^A(z_1)\hJ^B(z_2)=\frac{\hk/2}{(z_1-z_2)^2}\del^{AB}+
\frac{1}{z_1-z_2}if^{AB}_{\ \ \ \ C}\hJ^C(z_2)+{\rm r.t.} \nn \\
& &\la^A(z_1)\la^B(z_2)=\frac{\hk/2}{z_1-z_2}\del^{AB}+{\rm r.t.} \nn \\
& &\hJ^A(z_1)\la^B(z_2)=\frac{1}{z_1-z_2}if^{AB}_{\ \ \ \ C}\la^C
+{\rm r.t.}.
\ee
We have as usual taken $\hJ(z)=\hJ^A(z)t_A$ where $t_A$ are the generators
of the Lie algebra g of G and
$f^{AB}_{\ \ \ \ C}$ are its structure constants.

It is convenient to decouple the fermions from the affine currents, and
we hence introduce the new current $J^A\equiv \hJ^A+\frac{i}{\hk} f^A_{\ \
BC}\!:\!\la^B\la^C\!:$. (: : means as usual normal ordering.)
$J$ obeys an affine OPE with
level $\hk-c_g$ and the OPE between $J$ and $\la$ consists of
regular terms only.
The superconformal generators  may be realized
in terms of $\la^A$ and $J^A$ as
\be
T(z)=\frac{1}{\hk}\!:\!J_AJ^A\!:\!-\frac{1}{\hk}\!:\!\la_A\pr_z\la^A\!:, \
G(z)=\frac{2}{\hk}\!:\!J^A\la_A\!:\!-%
\frac{2i}{3\hk^2}f_{ABC}\!:\!\la^A\la^B\la^C\!:
\label{supconfdec}.
\ee
In this form it is easy to check that $j^A(Z)$ is a superconformal weight
one primary. 
\\

We now wish to consider the gauged supersymmetric WZNW model.
We take it to be gauged by a subgroup H, and
in a manifestly supersymmetric way. The
gauged action may be represented by two supersymmetric WZNW models,
the original
G valued supersymmetric WZNW model, and one valued over the subgroup H
\cite{Ts}.
We will refer to the latter as the auxiliary supersymmetric WZNW model or
shortly the auxiliary sector. We take the level of the auxiliary sector
to be $\hkt$.

To be more precise we have in terms of the partitionfunction
\be
Z=\int[{\rm d}{\cal G}][{\rm d}{\cal A}][{\rm d}\bar{\cal A}]
e^{S_{\hk}({\cal G},{\cal A},\bar{\cal A})}
=\int[{\rm d}{\cal G}][{\rm d}\widetilde{{\cal H}}]{\cal J}e^
{S_{\hk}({\cal G})+S_{\hkt}(\widetilde{{\cal H}})}\label{spartfunc}
\ee
where ${\cal A}$ and $\bar{\cal A}$ are the gauge superfields. ${\cal J}$
is the jacobian originating from a change of variables. It
contributes to the
conformal anomaly by a factor of $-3d_h$ where $d_h$ is the dimension
of the group H.
Although the effective action was provided by Tseytlin the BRST analysis was
not performed. 
\\

\section{BRST analysis}

The jacobian $\cal J$ in (\ref{spartfunc}) may be represented by a ghost
system consisting of fermionic ghosts $c,b$ of conformal weights (0,1) and
bosonic ghosts $\ga,\bet$ of weights $(\frac{1}{2},\frac{1}{2})$.
We take $c$ and $\ga$ to have ghost number one and $b$ and $\bet$ to have
minus one.
Introducing the super ghostfields $C(Z)=c(z)+\th\ga(z)$ and
$B(Z)=\bet(z)+\th b(z)$ of weights $(0,\frac{1}{2})$
and their antiholomorphic counterparts $\bar{C}(\bZ)=\bar{c}(\bz)
+\bth\bar{\ga}(\bz)$ and $\bar{B}(\bZ)=\bar{\bet}(\bz)+\bth \bar{b}(\bz)$
we can replace the jacobian by a ghost action
$S_{gh}=\int{\rm d}^2x{\rm d}^2\th\lef( B\bD C-\bar{B} D\bar{C}\rig).$

In superfields the BRST transformation of the effective action is given by
\be
& &\del_B{\cal G}=C{\cal G}-{\cal G}\bC, \
\del_B\widetilde{{\cal H}}=C\widetilde{{\cal H}}
-\widetilde{{\cal H}}\bC,\
\del_BC=\frac{1}{2}\{C,C\},\
\del_B\bar{C}=\frac{1}{2}\{\bar{C},\bar{C}\} \\
& &\del_BB=-\frac{\hk}{2\pi}D{\cal G}{\cal G}^{-1}-
\frac{\hkt}{2\pi}D\widetilde{{\cal H}}\widetilde{{\cal H}}^{-1}+[C,B],\
\del_B\bB=\frac{\hk}{2\pi}{\cal G}^{-1}
D{\cal G}+\frac{\hkt}{2\pi}\widetilde{{\cal H}}^{-1}D\widetilde{{\cal H}}
+[\bC,\bB]\nn. \label{SSSBRSTtrans}
\ee
which is a straightforward generalization of Bastianelli's result
for the bosonic case \cite{Ba}.
This transformation is trivially nilpotent for the ${\cal G}$, $C$, and
$\bC$ fields. For $B$ and $\bB$ we have non-trivial transformations
under $\del_B^2$ e.g. we find
$\del_B^2\bB=-\frac{\hk}{2\pi}{\cal G}^{-1}\bD C{\cal G}
-\frac{\hkt}{2\pi}\widetilde{{\cal H}}^{-1}\bD C\widetilde{{\cal
H}}+\frac{\hk+\hkt}{2\pi}\bD\bC.$
For nilpotency we thus need $\hk+\hkt=0$ and make use of the
holomorphisity of $C$. We get similar expressions for $\del_B^2B$.

We may expand (5) in components to find the
BRST transformation
\be
& &\del_Bg=cg-g\bc,\ \del_Bh=ch-h\bc, \
\del_Bc=\frac{1}{2}\{c,c\},\ \del_B\bc=\frac{1}{2}\{\bc,\bc\}  \\
& &\del_Bb=\frac{\hk}{2\pi}(\pr_zg\gi+\la\la)+
\frac{\hkt}{2\pi}(\pr_zh\hi+\tla\tla)+[\bet,\ga]+\{c,b\} \nn \\
& &\del_B\bb=\frac{\hk}{2\pi}(-\gi\pr_{\bz}g+\bla\bla)+
\frac{\hkt}{2\pi}(-\hi\pr_{\bz}h+\btla\btla)+[\bbet,\bga]+\{\bc,\bb\} \nn \\
& &\del_B\la=-\ga+\{c,\la\},\ \del_B\bla=\bga+\{\bc,\bla\},\
\del_B\tla=-\ga+\{c,\tla\},\ \del_B\btla=\bga+\{\bc,\btla\}\nn \\
& &\del_B\ga=[c,\ga],\ \del_B\bga=[\bc,\bga],\
\del_B\bet=-\frac{\hk}{2\pi}\la-\frac{\hkt}{2\pi}\tla+[c,\bet],\
\del_B\bbet=\frac{\hk}{2\pi}\bla+\frac{\hkt}{2\pi}\btla+[\bc,\bbet] \nn
\label{BRSTtrans}
\ee
{}From this transformation we may find the BRST charge.

First we for convenience
introduce the affine currents of the fermionic and the bosonic ghost
systems $J^a_{Fgh}(z)\equiv-if^{ab}_{\ \ c}\!:\!c_bb^c\!:\!(z)$ and
$J^a_{Bgh}(z)
\equiv if^{ab}_{\ \ c}\!:\!\ga_b\bet^c\!:\!(z)$.
They will satisfy OPE's
of affine type with levels $2c_h$ and $-2c_h$, respectively.
The index convention is that capital indices are g valued
and lower case h valued.

The BRST charge can now be written as
\be
& & Q\equiv \oint\frac{{\rm d}z}{2\pi}j_{BRST}(z)= \\
& &\oint\frac{{\rm d}z}{2\pi}\!:\!\lef(c_a(z)\lef(\hJ^a(z)+\hat{\;\ti{J}^a}(z)+
\frac{1}{2}J_{Fgh}^a(z)+J_{Bgh}^a(z)\rig)\!:\rig.+
\lef.\ga_a(z)\lef(\la^a(z)+\ti{\la}^a(z)\rig)\rig). \label{BRST}\nn
\ee
If we demand that $Q$ is nilpotent we find the restriction $\hk+\hkt=0$.

We may actually split the BRST charge (7) into two anticommuting and
nilpotent pieces. Introduce the gradation
${\rm grad}(\ga^a)=1,\ {\rm grad}(\bet^a)=-1. \label{gradation}$
This gives us $Q=Q_0+Q_1$ where
$Q_1=\oint\frac{{\rm d}z}{2\pi}\lef(\ga_a(z)(\la^a(z)+\ti{\la}^a(z))\rig).$
This degree is conserved by the OPE or equivalently
the commutator, hence $Q^2=0$ implies $Q_0^2=0$, $Q_1^2=0$ and $\{Q_0,Q_1\}=0$.
\\

\section{Superconformal algebra}

The total superconformal generators of the effective action becomes
\be
T^{tot}(z)&=&
\frac{1}{\hk}:\hJ^A\hJ_A:(z)-\frac{1}{\hk}:\la^A\pr_z\la_A:(z)+
\frac{1}{\hkt}:\hat{\;\ti{J}^a}\hat{\;\ti{J}_a}:(z)-
\frac{1}{\hkt}:\ti{\la}^a\pr_z\ti{\la}_a:(z) \nn \\
& &-:b^a\pr_zc_a(z):+
\frac{1}{2}:\pr_z\bet^a\ga_a:(z)-\frac{1}{2}:\bet^a\pr_z\ga_a:(z) \\
G^{tot}(z)&=&
\frac{2}{\hk}:\hJ^A\la_A:(z)-
\frac{2i}{3\hk^2}f_{ABC}:\la^A\la^B\la^C:(z)+\nn \\
& & \frac{2}{\hkt}:\hat{\;\ti{J}^A}\tla_A:(z)-
\frac{2i}{3\hkt^2}f_{ABC}:\tla^A\tla^B\tla^C:(z)+\pr_zc_a\bet^a(z)-
\ga_ab^a(z).
\ee
The conformal anomaly of this system, using the nilpotency restriction
$\hk+\hkt=0$, reads $c_{tot}=\frac{(\hk-c_g)d_g}{\hk}+\frac{1}{2}d_g-
\frac{(\hk-c_h)d_h}{\hk}
-\frac{1}{2}d_h$.
This may be recognized as the conformal anomaly of the algebraic N=1 coset
construction of Kazama and Suzuki \cite{KaSu}.

With the result of the
gauged bosonic WZNW model at hand \cite{KS}, this suggests that we
should look for a decomposition of $T^{tot}(z)$ and $G^{tot}(z)$
such that $T^{tot}(z)=T^{G/H}(z)+T^{ex}(z),\ G^{tot}(z)=G^{G/H}(z)+G^{ex}(z)$
where $T^{G/H}(z)\equiv T^G(z)-T^H(z)$ and $G^{G/H}(z)\equiv G^G(z)-G^H(z)$
are the superconformal generators of the Kazama-Suzuki coset construction.
$T^G(z)$ and $G^G(z)$ are given in eq.(\ref{supconfdec}), and
$T^H(z)$ and $G^H(z)$ are provided below.
Furthermore, we anticipate that $T^{ex}(z)$ and
$G^{ex}(z)$ are BRST exact.

Following \cite{KaSu}, we define $T^H(z)$ and $G^H(z)$ to be the generators
which transforms the subalgebra pieces $\hJ^a(z)$ and $\la^a(z)$
as components of
a weight one superfield. We must then define the new affine current
$J'^a(z)\equiv \hJ^a(z)+\frac{i}{\hk}f^a_{\ bc}:\la^b\la^c:(z)$
which is of level $\hk-c_h$.

We may now take the superconformal generators of the subalgebra to be
given by (\ref{supconfdec}) by interchangeing capital indices
for lower case indices and changeing $J$ into $J'$ \cite{KaSu}.
This provides an orthogonal decomposition in the desired sense i.e. the
OPE's
$T^{G/H}\hJ^a,\ T^{G/H}\la^a,\
G^{G/H}\hJ^a,\ G^{G/H}\la^a$ all gives only regular terms.

The BRST current $j_{BRST}(z)$ transforms under the superconformal algebra
as a conformal weight one field with a superpartner $u_{BRST}(z)$ of conformal
weight 1/2.
Explicitly we have $u_{BRST}(z)=-c_a(\la^a+\tla^a)(z)+
\frac{i}{2}f^{ab}_{\ \ c}:c_ac_b\bet^c:(z)$.
This means that we may construct the superconformal
weight one field $u_{BRST}(z)+\th j_{BRST}(z)$. The components satisfy OPE's
which are trivial in the sense that, $j_{BRST}u_{BRST}$ gives only regular
terms if and only if $\hk+\hkt=0$
and $u_{BRST}u_{BRST}$ always gives only regular terms.
Furthermore, we have that $u_{BRST}(z)$ is BRST exact, indeed
$u_{BRST}(w)=[Q,c_d\bet^d(w)]$.

The BRST current obeys identical OPE's with the superconformal generators
$T^{ex}$ and $G^{ex}$ as it does with $T^{tot}$ and $G^{tot}$. This
means that $T^{tot}$, $G^{tot}$, $T^{ex}$ and $G^{ex}$ are all BRST
invariant.

In order to find if $T^{ex}(z)$ is BRST exact it is convenient to work
with the commutator $\{Q,K(z)\}$ of some operator $K(z)$. It is
straightforward to verify that we indeed have
$T^{ex}(z)=\{Q,K^{ex}(z)\} \label{BRSTexact}$,
where $K^{ex}(z)$ is given by
\be
& &K^{ex}(z)= \\
& &
\frac{1}{\hk}b_a(\hJ^a-\hat{\;\ti{J}^a})(z)+\frac{1}{2\hk}\lef(\pr_z
\bet_a(\la^a-\tla^a)(z)
-\bet_a(\pr_z\la^a-\pr_z\tla^a)(z)\rig)+ \nn \\
& & \frac{4i}{3\hk^2}f_{abc}\lef(\bet^a(\la^b-\frac{1}{2}\tla^b)\hJ^c(z)+
\bet^a(\tla^b-\frac{1}{2}\la^b)\hat{\;\ti{J}^c}(z)+
b^a(\frac{1}{2}\la^b\la^c+\frac{1}{2}\tla^b\tla^c-\frac{1}{2}\la^b\tla^c)\rig)
\hs{10mm}\nn
\ee

The BRST exactness of the superpartner $G^{ex}(z)$ follows from the algebra and
the graded Jacobi identity. Obviously
$T^{ex}(z)$ and $G^{ex}(z)$ generates a
superconformal algebra (with vanishing central charge) as
may be seen from the fact that it corresponds to G/G gaugeing.
Consider now the OPE
$T^{ex}(z)G^{ex}(w)=\frac{3/2G^{ex}(w)}{(z-w)^2}+
\frac{\pr_wG^{ex}(w)}{z-w}+{\rm r.t.}.$
In terms of commutators this amounts to
$[L^{ex}_m,G^{ex}_r]=(\frac{1}{2}m-r)G^{ex}_{m+r}$
where we have used the modes $L^{ex}_m$ and $G^{ex}_r$ of the Laurent
expansion of $T^{ex}(z)$ and $G^{ex}(z)$ respectively. We have also
that $L^{ex}_m=\{Q,K_m^{ex}\}$ in terms of the modes
$K^{ex}_m$ of $K^{ex}(z)$. Putting this together, using the graded
Jacobi identity and BRST invariance of $G^{ex}(z)$ we may find
$(\frac{1}{2}m-r)G^{ex}_{m+r}=[L^{ex}_m,G^{ex}_r]=
[\{Q,K_m^{ex}\},G^{ex}_r]=-[\{K_m^{ex},G^{ex}_r\},Q].$
Since $r$ and $m$ are both arbitrary it is clear that $G(z)$ is BRST exact. \\

\section{Remarks}

There exists a somewhat different approach to the BRST analysis
of this model. We may, following Schnitzer \cite{Sc}, decouple the
fermions in the sense that we in the path-integral change variables from
$\psi_1$ and $\psi_2$ to $\la$ and $\bla$. This gives rise to an anomaly
which manifests itself as a shift in the level of the resulting bosonic
WZNW model $\hk\lra\hk-c_g$. We hence end up with free fermions and
the bosonic WZNW model with shifted level.

We use this procedure on the supersymmetric WZNW model gauged in the
manifestly supersymmetric way \cite{Ts}. What we find is two
bosonic WZNW models, the original one of level $\hk-c_g$ and the
auxiliary of level $\hkt-c_h$. The free fermions remains as above
and so does the ghost system.

The BRST charge decomposes into two mutually commuting pieces $Q=Q_B+Q_{\rm
ff}$
corresponding to the bosonic WZNW models together with the fermionic ghosts,
and the free fermion system together with the bosonic ghosts.
The BRST symmetry corresonding to the first part is well known \cite{KS,Ba}.
We find a part of the BRST charge $Q_B$ which squares to zero if
$\hk-c_g+\hkt-c_h
+2c_h=0$. The BRST charge of the free fermions $\la$ and $\tla$
remains as above except that we introduce an scaling
of $\tla$ such that we have nilpotency for the level restriction
$\hk+\hkt=c_g-c_h$.
This hence provides us with a different model than the one considered in
the main part of this paper. They coincide, however, for G=H. \\

{\large{\bf{Acknowledgments}}}
We would like to thank Stephen Hwang for suggesting this
project and for numerous discussions. Jens Lyng Petersen, Yu Ming
and J\o rgen Rasmussen of
the Niels Bohr Institute are also acknowledged for discussions
concerning this project. Jos$\acute{\rm e}$ Figueroa-O'Farrill and
Sonia Stanciu of Queen Mary and Westfield College
are acknowledged for discussions and for sharing their
results on similar issues prior to publication.


\end{document}